\documentclass[useAMS,usenatbib]{mn2e}
\usepackage{amsmath, amssymb}
\usepackage{natbib}
\usepackage{amssymb}
\usepackage{graphicx}
\usepackage{multirow}
\usepackage{times}
\usepackage[normalem]{ulem}
\usepackage{color}
\def\gsim{ \lower .75ex \hbox{$\sim$} \llap{\raise .27ex \hbox{$>$}} }
\def\lsim{ \lower .75ex\hbox{$\sim$} \llap{\raise .27ex \hbox{$<$}} }

\title{Observational Evidence for a Correlation Between Jet Power and
Black Hole Spin}
\author[R. Narayan  and J. E. McClintock]
{Ramesh Narayan$^1$\thanks{E-mail: rnarayan@cfa.harvard.edu (RN);
jmcclintock@cfa.harvard.edu (JEM)}; 
Jeffrey E. McClintock$^1$\\
$^1${Harvard-Smithsonian Center for Astrophysics, 60 Garden Street, Cambridge, MA 02138, USA}\\
}

\begin{document}
\date{\today}
\maketitle
\label{firstpage}

\begin{abstract}

We show that the 5-GHz radio flux of transient ballistic jets in black
hole binaries correlates with the dimensionless black hole spin
parameter $a_*$ estimated via the continuum-fitting method. The data
suggest that jet power scales either as the square of $a_*$ or the
square of the angular velocity of the horizon $\Omega_H$.  This is the
first direct evidence that jets may be powered by black hole spin energy.
The observed correlation validates the continuum-fitting method of
measuring spin.  In addition, for those black holes that have
well-sampled radio observations of ballistic jets, the correlation may
be used to obtain rough estimates of their spins.

\end{abstract}

\begin{keywords}
accretion, accretion discs -- black hole physics -- binaries: close
-- ISM: jets and outflows -- X-rays: binaries
\end{keywords}

\section{Introduction}

Accreting black holes, both supermassive and stellar mass, are
commonly observed to produce powerful relativistic jets
\citep{Zensus97,
Mirabel_Rodriguez99}.  Although
there now exists a wealth of data and many detailed models,
the mechanism that powers these jets remains a mystery.

The popular idea that jets are powered by the black hole (BH) 
goes back to the work of \citet{Penrose69}, who
showed that a spinning BH has free energy.  \citet{BZ77} proposed a
plausible mechanism whereby this free energy could be used to power an
astrophysical jet.  They suggested that magnetic fields in the
vicinity of an accreting BH would be twisted as a result of the
dragging of space-time by the rotating BH. The twisted field lines
will carry away energy from the BH, producing an electromagnetic jet.
The broad outlines of this model have been confirmed in numerical
simulations (e.g. \citealt{Tchekhovskoy+11}).

While a connection between BH spin energy and relativistic jets is
theoretically very appealing, there has been no direct observational
evidence for such a link. This is because, until recently, there was
no BH with a believable measurement of the dimensionless spin
parameter $a_* \equiv cJ/GM^2$, where $M$ and $J$ are the mass and
angular momentum of the BH.  The situation has now changed.  Methods
are now available --- one in particular, the continuum-fitting (CF)
method\footnote{A second method, based on modeling the relativistically
broadened X-ray iron K$\alpha$ line, is not considered in this Letter
(see Section 4 for a discussion).}
\citep{Zhang+97,Gierlinski+01,Shafee+06,Davis+06,McClintock+06} ---
that have enabled us to make plausibly reliable measurements of $a_*$
for several stellar mass BHs. With this sample of spin measurements,
we are now in a position to test whether jet power is related to BH
spin.  Such a test is the goal of this Letter.

In Section 2, we describe our sample of stellar mass BHs and collect
together the relevant observational data on BH spins and jet power. In
Section 3, we plot radio power against BH spin and demonstrate that there is
a significant correlation between the two quantities. We summarize and
discuss in Section 4.

\section{The data}

\subsection{BH sample and spin estimates}
\label{sample}

The CF method (see \citealt{McClintock+11} for a brief review) fits
the X-ray continuum spectrum of an accreting stellar mass BH using the
classic relativistic thin-disc model of \citet{Novikov_Thorne73}.
The spectral fit gives an estimate of the radius of the inner
edge of the accretion disc. The BH spin parameter $a_*$ is then
obtained by assuming that the disc edge is located at the innermost
stable circular orbit (ISCO) of the Kerr metric.  The CF method has
been developed in detail over the last several years and has been
shown to produce consistent results when multiple independent
observations of the same source are available (e.g.
\citealt{Steiner+09,Steiner+10}).  In addition, numerical simulations
have provided support for a crucial assumption of the model, namely
that the disc inner edge is close to the ISCO
(\citealt{Shafee+08,Penna+10,Kulkarni+11,Noble+11}).

\begin{table*}
\begin{center}
\begin{minipage}{\textwidth}
\caption{Parameters of transient BHBs in the sample}
\begin{center}
\begin{tabular}{|c|c|c|c|c|c|c|c|}

\hline
BH Binary & $a_*$ & $M~(M_\odot)$ & $D~({\rm kpc})$ & $i~({\rm deg})$ & $(S_{\nu})_{\rm max,5GHz}$ (Jy) & $S_0(\gamma=2)~({\rm Jy})$ & References \\ 
\hline
A0620--00 & $0.12\pm0.19$ & $6.61\pm0.25$ & $1.06\pm0.12$ & $51.0\pm0.9$ & 0.203 & 0.145 & 1, 6, 7 \\
XTE J1550--564 & $0.34\pm0.24$ & $9.10\pm0.61$ & $4.38\pm0.50$ & $74.7\pm3.8$ & 0.265 & 0.859 & 2, 6, 8 \\
GRO J1655--40 & $0.7\pm0.1$ & $6.30\pm0.27$ & $3.2\pm0.5$ & $70.2\pm1.9$ & 2.42 & 7.74 & 3, 4, 6, 9, 10 \\
GRS 1915+105 & $0.975\pm0.025$ & $14.0\pm4.4$ & $11.0\pm1.0$ & $66.0\pm2.0$ & 0.912 & 2.04 & 5, 6, 11, 12 \\
4U 1543--47 & $0.8\pm0.1$ & $9.4\pm1.0$ & $7.5\pm1.0$ & $20.7\pm1.5$ & $>1.16\times10^{-2}$ & $>4.31\times10^{-4}$ & 3, 6, 13 \\ 
\hline

\end{tabular}
\end{center}
\vspace{-0.25cm} References: (1) \citet{Gou+10}; (2)
\citet{Steiner+11}; (3) \citet{Shafee+06}; (4) \citet{Davis+06}; (5)
\citet{McClintock+06}; (6) \citet{Ozel+10}; (7) \citet{Kuulkers+99};
(8) \citet{Hannikainen+09}; (9) \citet{Hjellming_Rupen95}; (10)
\citet{Hannikainen+00}; (11) \citet{Rodriguez+95}; (12)
\citet{Fender+99}; (13) \citet{Park+04}.
\end{minipage}
\end{center}
\label{default}
\end{table*}

The spins of the BH primaries in nine BH binaries (BHBs) have been
measured using the CF method \citep{McClintock+11,Gou+11}.  Five of
these BHBs, namely, A0620--00, XTE J1550--564, GRO J1655--40, GRS
1915+105, 4U 1543--47, are transient systems
\citep{Remillard_McClintock06}.  These five systems have 
low-mass secondaries and undergo mass transfer via Roche lobe overflow.
They are of primary interest to us because during outburst, as they
approach the Eddington limit, they produce ballistic jets
(Section \ref{radio}).  The measured BH spin values $a_*$ and masses $M$,
along with distances $D$ and binary inclination angles $i$, are listed
in Table~1.  In the case of A0620--00 and XTE J1550--564, the error
estimates on the spins are taken from the original papers
\citep{Gou+10,Steiner+11}. The other three spins were measured in the
early days of the CF method \citep{Shafee+06,Davis+06,McClintock+06},
and we have doubled the published error estimates.

An additional four stellar mass BHs have spin estimates: LMC X-3
\citep{Davis+06}, M33 X-7 \citep{Liu+08,Liu+10}, LMC X-1
\citep{Gou+09} and Cyg X-1 \citep{Gou+11}. These are persistent BHBs
\citep{Remillard_McClintock06} which have high-mass companion stars
and undergo mass transfer via winds. Also, they do not show the kind
of transient behavior seen in the previous five objects and are
generally understood to belong to a different class. We ignore them in
this study.

\subsection{Jet radio power}
\label{radio}

\citet{Fender+04} identified a number of systematic properties in the
radio emission of BHB jets.  They showed that there are two kinds of
jets associated with specific spectral states of the X-ray source.
The first type of jet is observed in the hard spectral state as a
steady outflow. This jet is observable only out to a few tens of au
and is apparently not very relativistic.  The second and far more
dramatic jet, which is central to this Letter, is launched when a BHB
with a low-mass secondary undergoes a transient outburst
\citep{Fender+04}.  This powerful transient jet usually appears near
(or soon after) the time of outburst maximum, as the source switches
from its inital hard state to a soft state via the `steep power-law'
(SPL) state, a violently-variable state characterized by both strong
thermal and power-law components of emission
\citep{Remillard_McClintock06}.  Transient jets manifest themselves as
blobs of radio (and occasionally X-ray) emitting plasma that move
ballistically outward at relativistic speeds ($\gamma_{\rm jet}>2$).
Because these pc-scale jets resemble the kpc-scale jets seen in
quasars, BHBs that produce them are called microquasars
\citep{Mirabel_Rodriguez99}.

Ballistic jet ejection occurs at a very specific stage during the
spectral evolution of a given system \citep{Fender+04}.  As most
clearly demonstrated for the prototype microquasar GRS 1915+105
\citep{Fender_Belloni04}, this stage appears to correspond to the
inward-moving inner edge of the accretion disc reaching the ISCO,
which results in a shock or some other violent event that launches the
large-scale relativistic jet.  In this scenario, it appears reasonable
that the jet is launched within a few gravitational radii and hence
plausible that the spin energy of the BH could power the jet.  In
contrast, the steady jet in the hard state is thought to originate
much further out at $\sim 10-100$~GM/c$^2$ \citep{Markoff+05} where
the effects of spin are relatively weak.  Another virtue of the
ballistic jets for our purposes is that they occur at a sharply
defined luminosity (i.e. near Eddington) compared to the hard state
steady jets, which occur over a wide range of luminosity.  Ballistic
jets are thus better `standard candles'. For these reasons, in this
Letter we restrict our attention to ballistic jets from transient 
low-mass BHBs.

A typical ballistic jet blob is initially optically thick and has a
low radio power.  As the blob moves out and expands, the larger
surface area causes its radio power to increase. This continues until
the blob becomes optically thin, after which the flux declines
rapidly.
The overall behaviour
is generally consistent with an expanding conical jet (e.g.
\citealt{Hjellming_Johnston88}).

\begin{figure}
\begin{center}
\includegraphics[width=\columnwidth]{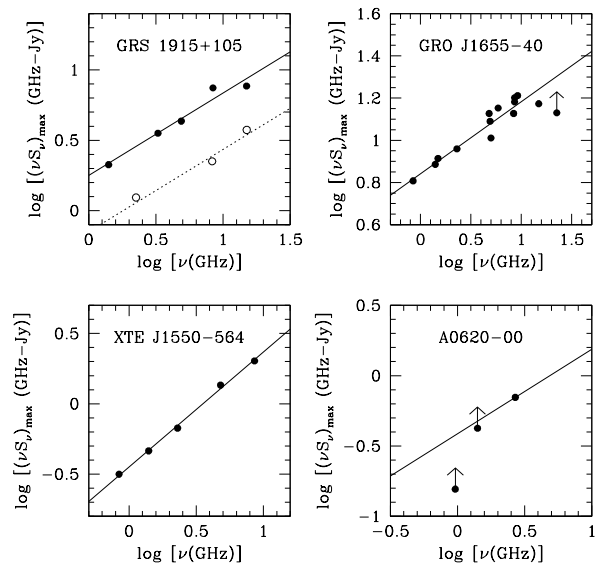}
\end{center}
\caption{Plot of the maximum observed radio power $(\nu S_\nu)_{\rm
    max}$ as a function of frequency $\nu$ for transient ballistic jet
  outbursts in four BHBs.  Two separate outbursts are shown for GRS
  1915+105. Best-fit lines (two separate ones in the case of GRS
  1915+105) are indicated, except in the case of A0620--00 where the
  line slope is fixed at 0.6 (or $\alpha=-0.4$).}
\label{fig:spectra}
\end{figure}

Fig. \ref{fig:spectra} shows the peak radio flux $(S_\nu)_{\rm max}$
versus $\nu$ observed at different radio frequencies $\nu$ for four of
the five transient BHBs in our sample. The radio light curves of these
four systems were monitored with good time resolution, allowing us to
obtain reasonably accurate estimates of the peak fluxes.  The top left
panel shows data for two separate outbursts of GRS 1915+105 (the solid
and open circles correspond respectively to the outbursts studied by
\citealt{Rodriguez+95} and \citealt{Fender+99}).  The two lines are
fits to the respective data and have a slope of 0.59; writing the
spectrum as $S_\nu \propto \nu^\alpha$, the fit corresponds to
$\alpha=-0.41$.  The top right panel combines the observations of
\citet{Hjellming_Rupen95} and \citet{Hannikainen+00} during an
outburst of GRO J1655--40.  The best-fit line corresponds to
$\alpha=-0.66$.\footnote{In the case of GRO J1655--40, the 22-GHz
observations did not cover the peak of the light curve. Hence this
point is shown as a lower limit.  Similarly, in A0620--00, the peak
was not observed at 0.962 and 1.14~GHz.}  The lower two panels
show data for XTE J1550--564 (\citealt{Hannikainen+09},
$\alpha=-0.18$) and A0620--00 \citep{Kuulkers+99}. For the latter
source, we do not have enough data points to determine the slope; the
line in the plot corresponds to $\alpha=-0.4$, the average spectral
index of the other three BHBs.  In order to enable a fair comparison
of the different objects, we use the fitted lines in the four panels
to estimate the peak fluxes $(S_\nu)_{\rm max}$ at a standard
frequency of 5~GHz. These 5-GHz peak flux values are listed in
Table~1.

While each of the above four objects was densely observed in radio
during one or more transient outbursts, 4U 1543--47 was unfortunately
not monitored well at radio frequencies during any of its several
outbursts.  The only radio data we know of when the source was bright
are those for the 2002 outburst summarized in \citet{Park+04}.  The
strongest radio flux was 0.022~Jy at 1.02675 GHz.  Assuming
$\alpha=-0.4$, this gives a flux of 0.0116~Jy at 5~GHz (or only
0.00043~Jy if one corrects for beaming with $\gamma_{\rm jet}=2$).  We
list this result separately in Table~1 and plot it as a lower limit in
Figs~2 and 3 because of the sparse radio coverage. In addition, there
was an anomaly in the 2002 X-ray outburst of this source.

The anomalous behaviour of 4U 1543--47 is apparent by an inspection of
figs~4--9 in \citet{Remillard_McClintock06}, which summarize in
detail the behaviour of six BH transients scrutinized by {\it RXTE}.
In panel $b$ of these figures, which displays light curves of the PCA
model flux coded by X-ray state, one sees that only 4U~1543--47 failed
to enter the SPL state (green triangles) near the peak of its
outburst, i.e. at the time of the radio coverage reported by Park et
al.  Rather, it remained locked in the thermal state (red crosses)
after its rise out of the hard state.  This behaviour contrasts sharply
with the behaviour of the other five transients which displayed the
strongly-Comptonized SPL state during both the late phase of their
rise to maximum and during their early decay phase.  Thus, because of
(1) the sparse radio coverage of 4U 1543--47, and (2) the failure of
the source to transition out of the jet-quenched thermal state
\citep{Gallo+03} to the SPL state (which is closely associated with
the launching of ballistic jets), we treat the maximum observed flux
of 0.022~Jy as a lower limit.  Finally, in sharp contrast to our
finding, we note that figs~5 and 6 in \citet{Fender+04} indicate a
very high jet power for 4U~1543--47.  We are unsure how they
arrived at their result, but suspect it was based on infrared data and
their equipartition model (see Section 4). If so, an extension of the present work to
infrared data might be worthwhile.

To measure jet power, we scale the 5-GHz peak flux of each BHB by the
square of the distance to the source $D$. We also divide by the BH
mass $M$ since we expect the power to be proportional to $M$ (this
scaling is not important since the range of masses is only a factor
$\sim2$). We thus obtain from the radio observations the following
quantity, which we treat as a proxy for the jet power:
\begin{equation}
P_{\rm jet} \equiv D^2 (\nu S_\nu)_{\rm max,5GHz}/M.
\end{equation}
It is hard to assess the uncertainty in the estimated values of
$P_{\rm jet}$. There is some uncertainty in the values of $D$ and $M$,
but these are not large. Potentially more serious, the radio flux may
not track jet power accurately.  For instance, the properties of the
ISM in the vicinity of the BHB may play a role and are likely to vary
from one object to another.  Also, the energy released in these
roughly Eddington-limited events will vary (e.g. see GRS 1915+105 in
Fig.~1).  Below, we arbitrarily assume that the uncertainty in $P_{\rm
jet}$ is 0.3 in the log, i.e. a factor of 2 each way.

\section{Jet power vs BH spin}
\label{correlation}

\begin{figure}
\begin{center}
\includegraphics[width=\columnwidth]{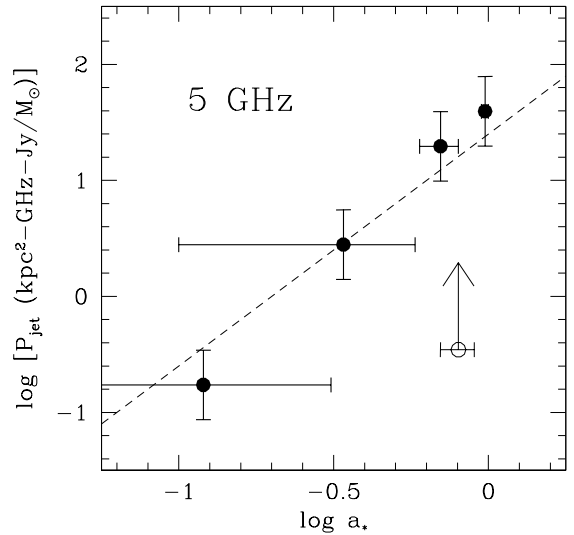}
\end{center}
\caption{Plot of the jet power $P_{\rm jet}$ as estimated from the
  maximum radio flux of ballistic jets (equation~1) vs the measured spin
  parameter of the BH $a_*$ for the transient BHBs in our
  sample. Solid circles correspond to the first four objects listed in
  Table 1, which have high quality radio data, and the open circle
  corresponds to 4U 1543--47, which has only a lower limit on the jet
  power. The dashed line corresponds to $P_{\rm jet} \propto a_*^2$,
  the theoretical scaling derived by \citet{BZ77}. The data suggest
  that ballistic jets derive their power from the spin of the central
  BH.}
\label{fig:astar5GHz.eps}
\end{figure}

Fig. 2 shows jet power $P_{\rm jet}$ plotted against BH spin
parameter $a_*$ for the five transient BHBs in our sample.  The data
are taken from Table~1.
The dashed line has a slope of 2, motivated by the
theoretical scaling, $P_{\rm jet} \propto a_*^2$, derived by
\citet{BZ77}.  The data points agree remarkably well with this
theoretical prediction.

\begin{figure}
\begin{center}
\includegraphics[width=\columnwidth]{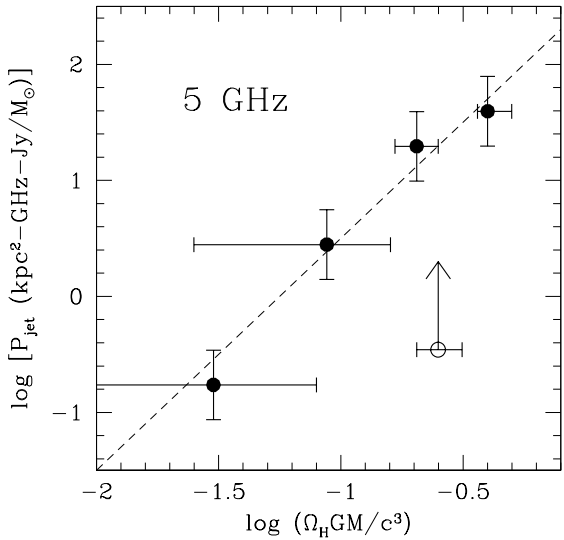}
\end{center}
\caption{Similar to Fig. 2, but showing the angular velocity of the BH
  horizon $\Omega_H$ along the abscissa. The dashed line
  corresponds to $P_{\rm jet}\propto \Omega_H^2$
  \citep{Tchekhovskoy+10}.  }
\label{fig:OmegaH5GHz.eps}
\end{figure}

\citet{BZ77} assumed a slowly spinning BH: $a_* \ll 1$.
\citet{Tchekhovskoy+10} obtained a more accurate theoretical scaling
which works up to spins fairly close to unity: $P_{\rm jet} \propto
\Omega_H^2$,
where $\Omega_H$ is the angular frequency of the
BH horizon, $\Omega_H = a_*(c^3/2GM)/(1+\sqrt{1-a_*^2})$.
Fig.~3 shows a plot of $P_{\rm jet}$ vs $\Omega_H$, with the dashed
line corresponding to a slope of 2.  The agreement is again very good.

We need to consider one additional effect: relativistic
beaming. Assuming a typical jet Lorentz factor $\gamma_{\rm jet}=2$
\citep{Fender+04} and using the inclination angles $i$ given in
Table~1, we have corrected the values of $(S_{\nu})_{{\rm max},5~{\rm
    GHz}}$.  The beaming-corrected radio fluxes $S_0$ (computed using
the relations given in \citealt{Mirabel_Rodriguez99} with the values
of $\alpha$ given in Section 2.2) are listed in Table 1.  The inferred jet
power of A0620--00 (the left-most point in Figs~2 and 3) decreases by
a small factor, whereas the other three jet powers increase by a
larger factor. As a result, the spread in $P_{\rm jet}$ among the four
objects becomes $\sim3.0$ orders of magnitude, compared to $\sim2.4$
orders in Figs~2, 3. Thus, allowing for beaming enhances the range of
$P_{\rm jet}$ in the sources and solidifies further the connection
between jet power and BH spin.

\section{Discussion}
\label{discussion}

Since the correlations shown in Figs~2 and 3 are based on only four
objects, one wonders whether we are merely seeing chance alignment of
intrinsically uncorrelated data.  The chief argument against this
hypothesis is that $a_*$ varies over the full allowable range of
prograde spins, $\Omega_H$ varies by more than a factor of 10, and
$P_{\rm jet}$ varies by 2.4 orders of magnitude (or 3 orders of
magnitude if one corrects for beaming assuming $\gamma_{\rm jet}=2$).
Also, the plotted points differ from one another by several standard
deviations, which is not statistically likely.  Therefore we conclude
that the power of ballistic jets is most likely correlated with the
spin of the accreting BH.

At the same time, and as a corollary, the strong apparent correlation
validates the CF method of measuring spin.  The CF method is based on
a number of assumptions, most of which have been independently
validated (see \citealt{McClintock+11}).  The results presented here
provide yet another validation. Caveats to the above conclusions include
the small size of the sample, insufficient data on one object (4U 1543-47),
and uncertainties in how well jet power and
radio luminosity track each other.

The existence of a correlation between jet power and BH spin does not
necessarily mean that the energy source for the jet is BH spin.  The
power could possibly be supplied by the accretion disc
\citep{Ghosh_Abramowicz97,Livio+99}.  Since the binding energy of a
particle at the ISCO increases with increasing $a_*$, the disc power
increases with BH spin and this might cause the observed correlation.
However, we note that the radiative luminosity of a thin accretion
disc varies by only a modest factor with BH spin; the radiative
efficiency $\eta=0.061$ for $a_*=0.12$ (the spin of A0620--00) and
$\eta = 0.23$ for $a_*=0.98$ (the spin of GRS 1915+105). If jet power
scales similarly, and if radio luminosity is roughly proportional to
jet power, we expect no more than a factor of 4 variation in
the radio powers in our sample. Instead, we see a factor
$\sim250-1000$.  Moreover, the observed spread is rather close to what
is expected theoretically
if jets are powered by BH spin. The evidence thus suggests that
ballistic jets are powered directly by the spin energy of the
accreting BH.

Based on the above arguments, we view our results as a confirmation of
the Penrose-Blandford-Znajek mechanism of powering relativistic jets
by BH spin energy.  Theoretically, this mechanism
depends on both the BH spin and the magnetic field strength at the
horizon.  The latter is believed to depend on the mass accretion rate
$\dot{M}$ (e.g.
\citealt{Tchekhovskoy+11}).  Since
ballistic jets are seen during a very specific phase of the evolution
of a transient BHB, it is reasonable to assume that $\dot{M}$
(normalized by the BH mass) is roughly the same in different objects
when they exhibit ballistic jets, or indeed in different ballistic jet
episodes in the same object.  This allows us to eliminate $\dot{M}$
from our analysis and to treat ballistic jets as `standard candles', 
thereby making the comparisons shown in Figs~2 and 3
meaningful.

In addition to the CF method, a second method based on fitting the
profile of the relativistically broadened Fe K$\alpha$ line has been
used to estimate BH spins \citep{Reynolds_Nowak03,Miller07}.  In this
Letter, we opt to use only CF spin data for two reasons: (1) The
Fe-line models are complex and therefore relatively less reliable. CF
spins are obtained rather simply by modeling a dominant thermal disc
component, while Fe-line spins require modeling the thermal disc plus
a Compton component plus a disc reflection component, which includes
the Fe K$\alpha$ line.  The Fe-line method furthermore requires
characterizing a luminous corona of unknown geometry. (2) For several
well-studied systems, the Fe-line method has generated widely
inconsistent values of the spin parameter or shown to be strongly
model dependent (e.g. for Cyg X-1, see Sec. 7.1 in \citealt{Gou+11};
for MCG--6-30-15, see \citealt{Miller+09}).  The CF method, on the
other hand, gives consistent results for multiple and independent
observations of individual sources.  For example, for the BHBs listed
in Table 1 (excluding A0620--00), consistent results were obtained for
$\approx50$ {\it RXTE} spectra (XTE J1550--564); 2 {\it ASCA} and 31
{\it RXTE} spectra (GRO J1655--40); 1 {\it ASCA} and 5 {\it RXTE}
spectra (GRS 1915+105); and 34 {\it RXTE} spectra (4U 1543--47).  The
standout example is LMC X-3 with 411 spectra collected by eight X-ray
missions over 26 years \citep{Steiner+10}.

After considering separately both ballistic and hard state steady
jets, \citet{Fender+10} find no evidence for a correlation between jet
power and BH spin.  We have already given in Section 2.2 a plausible reason
for the absence of evidence in the case of the steady jets.  We now
focus on rationalizing the very different results obtained by Fender
et al. (presented in their Section 2.2.1 and fig.~6) and ourselves for the
ballistic jets.  Our data sample (Table~1) is identical to their comparable
sample (see the right panel of their fig.~6).  The only significant
difference in data selection is that for GRO J1550--564 we use the new
spin value of \citet{Steiner+11}, $a_*=0.34 \pm 0.24$, while they used
the earlier \citet{Davis+06} limit of $a_*<0.8$.

The substantial difference between our results and those of
\citet{Fender+10} is, in the end, determined by the choice of the
quantity used to represent jet power.  We simply use the maximum observed flux
density at 5 GHz expressed as a luminosity.  Fender et al.  compute
jet power from the peak radio luminosity and the rise time of some
particular synchrotron event.  The authors clearly state that their
approach `is useful to provide lower limits on, and
order-of-magnitude estimates of, jet power but is very susceptible to
errors resulting from poor sampling of events, uncertainties in
Doppler boosting, assumptions about equipartition, etc.'  Their
estimates of jet luminosity for three sources are given in Table~1 of
\citet{Fender+04}, but it is not clear how the luminosities of
A0620--00 and 4U~1543--47 were estimated.  The authors further adopt a
formula relating jet power to X-ray luminosity, log$_{\rm 10}L_{\rm
  jet} = c~+~0.5($log$_{\rm 10}L_{\rm x}~-~34)$, and estimate the
normalization constant $c$ in the preceding formula, which they treat
as their proxy for jet power.  In short, their proxy for jet power is
heavily model dependent and ours is model independent.

The correlation shown in Fig.~2 can be used to obtain rough estimates
of spin for any transient BH that has undergone a major outburst cycle
and that has been closely monitored at radio wavelengths. For
instance, radio observations of Nova Muscae 1991 (GRS 1124--68) by
\citet{Ball+95} suggest a maximum 5-GHz radio flux $\approx0.2$~Jy.
Assuming a distance $D\approx6$~kpc and a typical BH
mass $M\approx8M_\odot$ \citep{Ozel+10}, we obtain $\log[P_{\rm
    jet}]\approx 0.65\pm0.3$.  Fig.~2 then gives $a_* \approx0.3-0.6$.
In the case of GX 339-4, the brightest X-ray and radio outburst
\citep{Gallo+04} had a maximum 5-GHz flux of 0.055~Jy. Taking
$D\approx 9$~kpc, $M\approx 8M_\odot$ \citep{Ozel+10}, we find
$\log[P_{\rm jet}]\approx 0.4\pm0.3$ and $a_* \approx 0.2-0.5$.  The latter
estimate is consistent with the strict upper limit $a_*<0.9$ derived
by \citet{Kolehmainen_Done10} using the CF method with conservative
assumptions.

These examples illustrate the importance of obtaining good radio
coverage for all future transient BHBs, including especially the
recurrent system 4U 1543--47.  Those systems that have CF-based spin
measurements will flesh out the correlations plotted in Figs~2 and 3.
For the many other BH transients that lack a sufficiently-bright
optical counterpart and are therefore out of reach of the CF method,
the radio data can either be used as a check on Fe-line measurements
of spin or serve as our only estimate of spin.

Acknowledgements: The authors thank Jack Steiner for useful
discussions.  This work was supported in part by NASA grants NNX11AE16G
(to RN) and NNX11AD08G (to JEM).

\bibliography{ms}

\begin{thebibliography}{48}
\expandafter\ifx\csname natexlab\endcsname\relax\def\natexlab#1{#1}\fi

\bibitem[{Ball} et~al.(1995){Ball}, {Kesteven}, {Campbell-Wilson}, {Turtle} \&
  {Hjellming}]{Ball+95}
{Ball} L., {Kesteven} M.~J., {Campbell-Wilson} D., {Turtle} A.~J., {Hjellming}
  R.~M., 1995, \mnras, 273, 722

\bibitem[{Blandford} \& {Znajek}(1977)]{BZ77}
{Blandford} R.~D., {Znajek} R.~L., 1977, \mnras, 179, 433

\bibitem[{Davis} et~al.(2006){Davis}, {Done} \& {Blaes}]{Davis+06}
{Davis} S.~W., {Done} C., {Blaes} O.~M., 2006, \apj, 647, 525

\bibitem[{Fender} \& {Belloni}(2004)]{Fender_Belloni04}
{Fender} R., {Belloni} T., 2004, \araa, 42, 317

\bibitem[{Fender} et~al.(2004){Fender}, {Belloni} \& {Gallo}]{Fender+04}
{Fender} R.~P., {Belloni} T.~M., {Gallo} E., 2004, \mnras, 355, 1105

\bibitem[{Fender} et~al.(2010){Fender}, {Gallo} \& {Russell}]{Fender+10}
{Fender} R.~P., {Gallo} E., {Russell} D., 2010, \mnras, 406, 1425

\bibitem[{Fender} et~al.(1999){Fender}, {Garrington}, {McKay}
  et~al.]{Fender+99}
{Fender} R.~P., {Garrington} S.~T., {McKay} D.~J., et~al., 1999, \mnras, 304,
  865

\bibitem[{Gallo} et~al.(2004){Gallo}, {Corbel}, {Fender}, {Maccarone} \&
  {Tzioumis}]{Gallo+04}
{Gallo} E., {Corbel} S., {Fender} R.~P., {Maccarone} T.~J., {Tzioumis} A.~K.,
  2004, \mnras, 347, L52

\bibitem[{Gallo} et~al.(2003){Gallo}, {Fender} \& {Pooley}]{Gallo+03}
{Gallo} E., {Fender} R.~P., {Pooley} G.~G., 2003, \mnras, 344, 60

\bibitem[{Ghosh} \& {Abramowicz}(1997)]{Ghosh_Abramowicz97}
{Ghosh} P., {Abramowicz} M.~A., 1997, \mnras, 292, 887

\bibitem[{Gierli{\'n}ski} et~al.(2001){Gierli{\'n}ski},
  {Macio{\l}ek-Nied{\'z}wiecki} \& {Ebisawa}]{Gierlinski+01}
{Gierli{\'n}ski} M., {Macio{\l}ek-Nied{\'z}wiecki} A., {Ebisawa} K., 2001,
  \mnras, 325, 1253

\bibitem[{Gou} et~al.(2009){Gou}, {McClintock}, {Liu} et~al.]{Gou+09}
{Gou} L., {McClintock} J.~E., {Liu} J., et~al., 2009, \apj, 701, 1076

\bibitem[{Gou} et~al.(2011){Gou}, {McClintock}, {Reid} et~al.]{Gou+11}
{Gou} L., {McClintock} J.~E., {Reid} M.~J., et~al., 2011, ApJ, in press, ArXiv
  e-prints, arXiv: 1106.3690 [astro-ph.HE]

\bibitem[{Gou} et~al.(2010){Gou}, {McClintock}, {Steiner} et~al.]{Gou+10}
{Gou} L., {McClintock} J.~E., {Steiner} J.~F., et~al., 2010, \apjl, 718, L122

\bibitem[{Hannikainen} et~al.(2000){Hannikainen}, {Hunstead}, {Campbell-Wilson}
  et~al.]{Hannikainen+00}
{Hannikainen} D.~C., {Hunstead} R.~W., {Campbell-Wilson} D., et~al., 2000,
  \apj, 540, 521

\bibitem[{Hannikainen} et~al.(2009){Hannikainen}, {Hunstead}, {Wu}
  et~al.]{Hannikainen+09}
{Hannikainen} D.~C., {Hunstead} R.~W., {Wu} K., et~al., 2009, \mnras, 397, 569

\bibitem[{Hjellming} \& {Johnston}(1988)]{Hjellming_Johnston88}
{Hjellming} R.~M., {Johnston} K.~J., 1988, \apj, 328, 600

\bibitem[{Hjellming} \& {Rupen}(1995)]{Hjellming_Rupen95}
{Hjellming} R.~M., {Rupen} M.~P., 1995, Nature, 375, 464

\bibitem[{Kolehmainen} \& {Done}(2010)]{Kolehmainen_Done10}
{Kolehmainen} M., {Done} C., 2010, \mnras, 406, 2206

\bibitem[{Kulkarni} et~al.(2011){Kulkarni}, {Penna}, {Shcherbakov}
  et~al.]{Kulkarni+11}
{Kulkarni} A.~K., {Penna} R.~F., {Shcherbakov} R.~V., et~al., 2011, \mnras,
  414, 1183

\bibitem[{Kuulkers} et~al.(1999){Kuulkers}, {Fender}, {Spencer}, {Davis} \&
  {Morison}]{Kuulkers+99}
{Kuulkers} E., {Fender} R.~P., {Spencer} R.~E., {Davis} R.~J., {Morison} I.,
  1999, \mnras, 306, 919

\bibitem[{Liu} et~al.(2008){Liu}, {McClintock}, {Narayan}, {Davis} \&
  {Orosz}]{Liu+08}
{Liu} J., {McClintock} J.~E., {Narayan} R., {Davis} S.~W., {Orosz} J.~A., 2008,
  \apjl, 679, L37

\bibitem[{Liu} et~al.(2010){Liu}, {McClintock}, {Narayan}, {Davis} \&
  {Orosz}]{Liu+10}
{Liu} J., {McClintock} J.~E., {Narayan} R., {Davis} S.~W., {Orosz} J.~A., 2010,
  \apjl, 719, L109

\bibitem[{Livio} et~al.(1999){Livio}, {Ogilvie} \& {Pringle}]{Livio+99}
{Livio} M., {Ogilvie} G.~I., {Pringle} J.~E., 1999, \apj, 512, 100

\bibitem[{Markoff} et~al.(2005){Markoff}, {Nowak} \& {Wilms}]{Markoff+05}
{Markoff} S., {Nowak} M.~A., {Wilms} J., 2005, \apj, 635, 1203

\bibitem[{McClintock} et~al.(2011){McClintock}, {Narayan}, {Davis}
  et~al.]{McClintock+11}
{McClintock} J.~E., {Narayan} R., {Davis} S.~W., et~al., 2011, Classical and
  Quantum Gravity, 28, 11, 114009

\bibitem[{McClintock} et~al.(2006){McClintock}, {Shafee}, {Narayan},
  {Remillard}, {Davis} \& {Li}]{McClintock+06}
{McClintock} J.~E., {Shafee} R., {Narayan} R., {Remillard} R.~A., {Davis}
  S.~W., {Li} L.-X., 2006, \apj, 652, 518

\bibitem[{Miller}(2007)]{Miller07}
{Miller} J.~M., 2007, \araa, 45, 441

\bibitem[{Miller} et~al.(2009){Miller}, {Turner} \& {Reeves}]{Miller+09}
{Miller} L., {Turner} T.~J., {Reeves} J.~N., 2009, \mnras, 399, L69

\bibitem[{Mirabel} \& {Rodr{\'{\i}}guez}(1999)]{Mirabel_Rodriguez99}
{Mirabel} I.~F., {Rodr{\'{\i}}guez} L.~F., 1999, \araa, 37, 409

\bibitem[{Noble} et~al.(2011){Noble}, {Krolik}, {Schnittman} \&
  {Hawley}]{Noble+11}
{Noble} S.~C., {Krolik} J.~H., {Schnittman} J.~D., {Hawley} J.~F., 2011, ArXiv
  e-prints, arXiv: 1105.2825 [astro-ph.HE]

\bibitem[{Novikov} \& {Thorne}(1973)]{Novikov_Thorne73}
{Novikov} I.~D., {Thorne} K.~S., 1973, in { Black Holes (Les Astres Occlus).
  Gordon and Breach, Paris\/}, edited by {C.~Dewitt \& B.~S.~Dewitt},  343--450

\bibitem[{{\"O}zel} et~al.(2010){{\"O}zel}, {Psaltis}, {Narayan} \&
  {McClintock}]{Ozel+10}
{{\"O}zel} F., {Psaltis} D., {Narayan} R., {McClintock} J.~E., 2010, \apj, 725,
  1918

\bibitem[{Park} et~al.(2004){Park}, {Miller}, {McClintock} et~al.]{Park+04}
{Park} S.~Q., {Miller} J.~M., {McClintock} J.~E., et~al., 2004, \apj, 610, 378

\bibitem[{Penna} et~al.(2010){Penna}, {McKinney}, {Narayan}, {Tchekhovskoy},
  {Shafee} \& {McClintock}]{Penna+10}
{Penna} R.~F., {McKinney} J.~C., {Narayan} R., {Tchekhovskoy} A., {Shafee} R.,
  {McClintock} J.~E., 2010, \mnras, 408, 752

\bibitem[{Penrose}(1969)]{Penrose69}
{Penrose} R., 1969, Nuovo Cimento Rivista Serie, 1, 252

\bibitem[{Remillard} \& {McClintock}(2006)]{Remillard_McClintock06}
{Remillard} R.~A., {McClintock} J.~E., 2006, \araa, 44, 49

\bibitem[{Reynolds} \& {Nowak}(2003)]{Reynolds_Nowak03}
{Reynolds} C.~S., {Nowak} M.~A., 2003, \physrep, 377, 389

\bibitem[{Rodriguez} et~al.(1995){Rodriguez}, {Gerard}, {Mirabel}, {Gomez} \&
  {Velazquez}]{Rodriguez+95}
{Rodriguez} L.~F., {Gerard} E., {Mirabel} I.~F., {Gomez} Y., {Velazquez} A.,
  1995, ApJS, 101, 173

\bibitem[{Shafee} et~al.(2006){Shafee}, {McClintock}, {Narayan}, {Davis}, {Li}
  \& {Remillard}]{Shafee+06}
{Shafee} R., {McClintock} J.~E., {Narayan} R., {Davis} S.~W., {Li} L.-X.,
  {Remillard} R.~A., 2006, \apjl, 636, L113

\bibitem[{Shafee} et~al.(2008){Shafee}, {McKinney}, {Narayan}, {Tchekhovskoy},
  {Gammie} \& {McClintock}]{Shafee+08}
{Shafee} R., {McKinney} J.~C., {Narayan} R., {Tchekhovskoy} A., {Gammie} C.~F.,
  {McClintock} J.~E., 2008, \apjl, 687, L25

\bibitem[{Steiner} et~al.(2010){Steiner}, {McClintock}, {Remillard}, {Gou},
  {Yamada} \& {Narayan}]{Steiner+10}
{Steiner} J.~F., {McClintock} J.~E., {Remillard} R.~A., {Gou} L., {Yamada} S.,
  {Narayan} R., 2010, \apjl, 718, L117

\bibitem[{Steiner} et~al.(2009){Steiner}, {McClintock}, {Remillard}, {Narayan}
  \& {Gou}]{Steiner+09}
{Steiner} J.~F., {McClintock} J.~E., {Remillard} R.~A., {Narayan} R., {Gou} L.,
  2009, \apjl, 701, L83

\bibitem[{Steiner} et~al.(2011){Steiner}, {Reis}, {McClintock}
  et~al.]{Steiner+11}
{Steiner} J.~F., {Reis} R.~C., {McClintock} J.~E., et~al., 2011, \mnras, 416,
  941

\bibitem[{Tchekhovskoy} et~al.(2010){Tchekhovskoy}, {Narayan} \&
  {McKinney}]{Tchekhovskoy+10}
{Tchekhovskoy} A., {Narayan} R., {McKinney} J.~C., 2010, \apj, 711, 50

\bibitem[{Tchekhovskoy} et~al.(2011){Tchekhovskoy}, {Narayan} \&
  {McKinney}]{Tchekhovskoy+11}
{Tchekhovskoy} A., {Narayan} R., {McKinney} J.~C., 2011, MNRAS, 418, L79

\bibitem[{Zensus}(1997)]{Zensus97}
{Zensus} J.~A., 1997, \araa, 35, 607

\bibitem[{Zhang} et~al.(1997){Zhang}, {Cui} \& {Chen}]{Zhang+97}
{Zhang} S.~N., {Cui} W., {Chen} W., 1997, \apjl, 482, L155

\end{thebibliography}

\end{document}